\documentclass[a4paper, fleqn, leqno, 11p]{article}
\usepackage{color,amsmath,amssymb,soul,graphicx}
\usepackage[footnotesize]{caption}
\usepackage[comma, square, numbers, sort&compress]{natbib} 

\begin{document}

\title{Cellular function given parametric variation: excitability in the Hodgkin-Huxley model}

\author{Hillel Ori\thanks{Technion -- Israel Institute of Technology, Haifa 32000, Israel}~, Eve Marder\thanks{Brandeis University, Waltham, MA 02453-2728, US}~ \& Shimon Marom\footnotemark[1]}

\maketitle

\begin{abstract}
\noindent How is reliable physiological function maintained in cells despite considerable variability in the values of key parameters of multiple interacting processes that govern that function?  Here we use the classic Hodgkin-Huxley formulation of the squid giant axon action potential to propose a possible approach to this problem. Although the full Hodgkin-Huxley model is very sensitive to fluctuations that independently occur in its many parameters, the outcome is in fact determined by simple combinations of these parameters along two physiological dimensions: Structural and Kinetic (denoted \textit{S} and \textit{K}). Structural parameters describe the properties of the cell, including its capacitance and the densities of its ion channels.  Kinetic parameters are those that describe the opening and closing of the voltage-dependent conductances.  The impacts of parametric fluctuations on the dynamics of the system -- seemingly complex in the high dimensional representation of the Hodgkin-Huxley model -- are tractable when examined within the \textit{S--K} plane. We demonstrate that slow inactivation, a ubiquitous activity-dependent feature of ionic channels, is a powerful local homeostatic control mechanism that stabilizes excitability amid changes in structural and kinetic parameters. 
\end{abstract}

\noindent The canonical Hodgkin-Huxley mathematical model of membrane excitability is embedded in a high dimensional parameter space. In their original report \cite{Hodgkin1952} Hodgkin and Huxley indicate that the parameters vary substantially between different cells, an observation that is extensively documented by electrophysiologists who have studied excitable membranes over half a century \cite{Johnstone2016,OLeary2015}. This cell-to-cell intrinsic variability -- i.e., variability that cannot be attributed to measurement uncertainties \cite{Johnstone2016,mirams2016uncertainty} -- is habitually averaged out in attempts to generalize findings \cite{Golowasch2002,Marder2011}. 

There are many reasons to assume that cell-to-cell variability is also expressed in a given cell over time \cite{Brenner2015,AsafGal12012010,Li2011,Marder2014,OLeary2014,Raj2008,Sigal2006}. For instance: kinetic parameters of channel gating change due to continuous modulation and activity-dependent roaming in protein configuration space; the ratio between the number of different channel proteins in the membrane might change due to differential protein expression or turnover; the membrane capacitance and leak conductance change during massive cell growth, movement or contact of the cell with biological matrices that impact on membrane surface tension. 

Randomly and independently pulled from the physiological range indicated by Hodgkin and Huxley, many combinations of parameters give rise to an excitable `solution' (i.e., stimulus-driven excitable membranes that generate single action potentials), but many other combinations lead to either non-excitable or oscillatory, pacemaking membranes \cite{Marom2016}. The solutions are sensitive to relatively slight parametric independent variations. This is in contrast to the biological neurons and muscles that maintain relatively invariant patterns of activity that are seemingly more robust when compared to the classic Hodgkin-Huxley type  models used to describe them. Cell-to-cell and within-cell parametric variation challenge our understanding of establishment and maintenance of excitability, as well as the methods we use in order to extract parameters from voltage-clamp data and construct suitable models \cite{Daly2015,Johnstone2016,sarkar2010regression,sobie2009parameter}. This essential problem goes beyond the regulation of excitability; it belongs to a class of open questions that concern the study of organization in biological systems and the emergence of macroscopic functional order from a large space of potential microscopic `disordered' configurations \cite{Braun2015}.

\begin{figure}[]
\centering
\includegraphics[width=12cm,height=9cm]{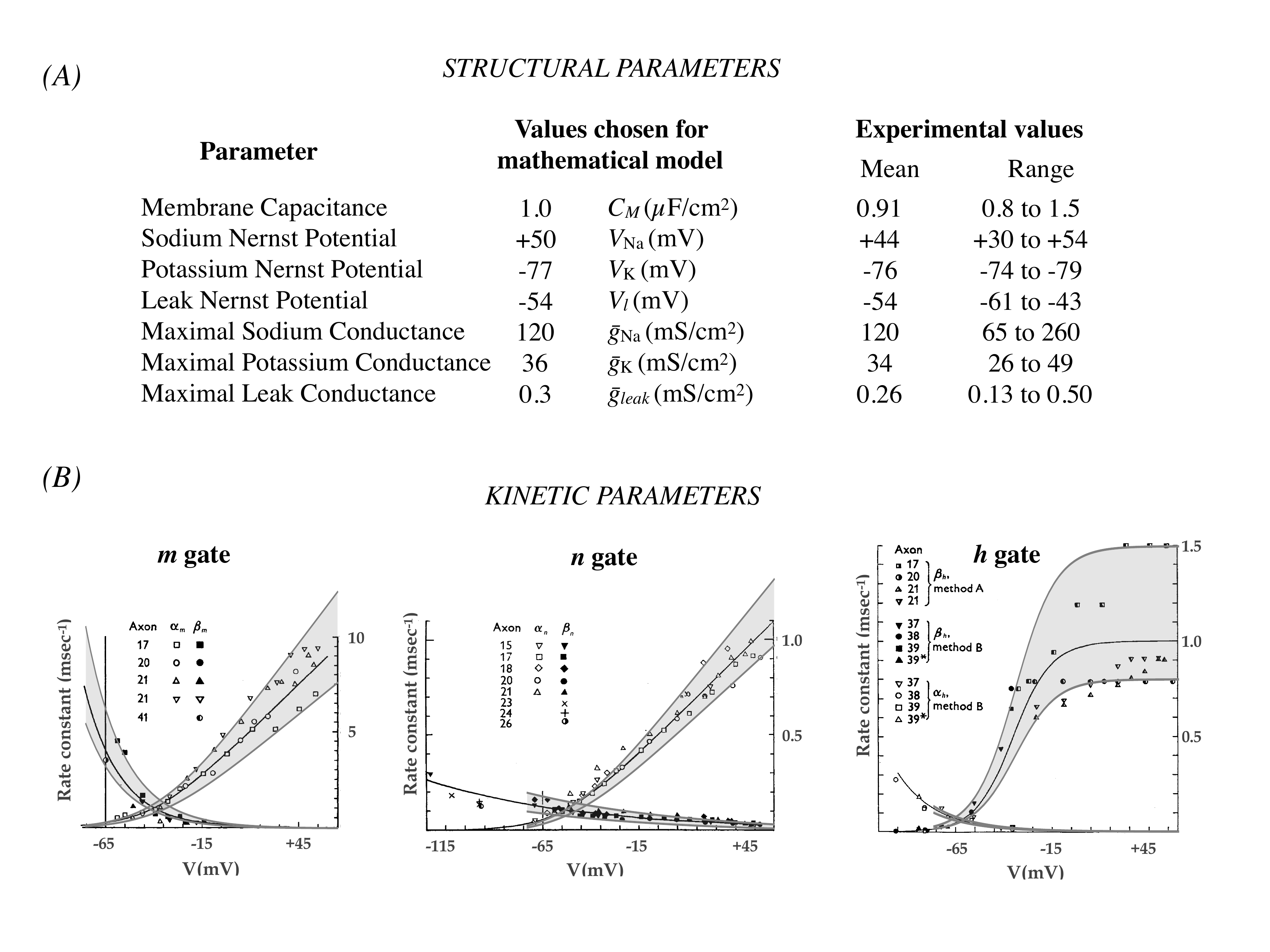}
\caption{\protect\rule{0ex}{20ex}a collage of slightly modified images extracted from the 1952 original report of Hodgkin and Huxley. The table (top panel) indicates ranges of cellular-level structural parameters. The graphs of the bottom panel depict protein-level kinetic parameters, expressed as six transition rate functions superposed with data points. The shaded areas added to the graphs of the bottom panel suggest that a linear scaling of transition rate functions captures most of the underlying variance: $<\alpha_{n}(v)> = [0.85, 1.15]$ , $<\beta_{n}(v)> = [0.7, 1.3]$, $<\alpha_{m}(v)> = [0.8, 1.3]$, $<\beta_{m}(v)> = [0.7, 1.35]$, $<\alpha_{h}(v)>= [0.85, 1.15]$, $<\beta_{h}(v)>= [0.8, 1.5]$.}
\label{fig:1}
\end{figure}

Attempts to account for invariant excitability given parametric variation focus on activity-dependent, homeostatic coordination of channel protein expression \cite{lemasson1993activity,liu1998model,OLeary2014,O_Leary_2018}. The interpretation is corroborated by correlations between mRNA concentrations of different ionic channel proteins \cite{Schulz2007}. It is also supported by elegant simulations showing how centralized activity-dependent (feedback) regulation that controls protein expression may navigate a cell into one of many functional solutions \cite{OLeary2014,OLeary2013}. But channel protein densities are not the only determinants of membrane excitability status. Even in the relatively simple Hodgkin-Huxley model -- a single compartment with two voltage-dependent conductances -- more than ten parameters are involved, possibly varying and impacting each other in a wide range of time scales (sub-second to hours and days). 

We acknowledge the difference between the high-dimensional parameter space dictated by the explicit Hodgkin-Huxley model, and the dimensionality of the physiological space within which regulation of excitability is embedded. In other words, we ask how many physiologically-relevant dimensions are needed to capture the dynamics of excitability and its regulation in the Hodgkin-Huxley formulation, as this may be very different from the number of free parameters in the full Hodgkin-Huxley model. From the physiologist's perspective, the actual (hopefully not too many) dimensions should be expressed in parameters that can be directly extracted from standard voltage-clamp data. The approach we take below maintains the biophysically measurable parameters and is therefore different from most other reductions previously done.  

We show that -- congruent with low dimensional models of excitability \cite{Abbott1990,FitzHugh1961,Izhikevich2003,Jack1975} -- the phenomenon of excitability may be reduced to two-dimensions. We identify these dimensions as cellular-level structural (denoted \textit{S}) and protein-intrinsic kinetic (denoted \textit{K}) dimensions and express them as combinations of actual Hodgkin-Huxley parameters that may be extracted from voltage-clamp data.  Structural parameters refer to membrane surface area, number of different voltage-dependent conductances, ionic concentrations.  Kinetic parameters are those that describe the rate constants associated with channel opening and closing.  Reduced to the \textit{S--K} space, the manifold of functional solutions is simpler to understand amidst parametric variations, enabling regulation of excitability by one activity-dependent principle that is tightly related to a ubiquitous physiological process: slow inactivation of ionic channels.

\subsection*{Results \& Discussion}

\subsubsection*{Hodgkin-Huxley model, multiple solutions and their sensitivity to parametric variations} 

The dispersion of parameters measured by Hodgkin and Huxley is summarized in Figure~\ref{fig:1} -- a collage of data and images from the original report, adapted to modern notation. The table (top panel) indicates ranges of cellular-level structural parameters. The term `structural' is used as these parameters are fully determined by physical measures that are characteristic of the cell. These include membrane surface area, the number of voltage-dependent conductances, and relevant equilibrium potentials. The graphs of Figure~\ref{fig:1} depict protein-level kinetic parameters, expressed as six transition rate functions superposed with data points. The mathematical expressions of these six rate functions -- each of which describes the change of transition rate with membrane voltage ($V$) -- involve more than ten different `hidden' parameters. Note the dispersion of points around the fitted rate functions, depicting repeated measurements in different axons. 

To simplify matters, the following analysis ignores variations in equilibrium potentials and focuses on ten parameters: membrane capacitance ($C_{\text{m}}$), maximal sodium, potassium and leak conductance ($\bar{g}_{\text{Na}}$, $\bar{g}_{\text{K}}$ and $\bar{g}_{\text{leak}}$, respectively), and the six transition rates underlying the opening and closing of `gates' -- $\alpha_{m}(v)$, $\beta_{m}(v)$, $\alpha_{h}(v)$, $\beta_{h}(v)$, $\alpha_{n}(v)$ and $\beta_{n}(v)$ -- as explained below. 

We begin creating a straw man, considering marginal (i.e. independent) uniformly distributed variations for all ten parameters. Values of parameters are expressed in terms of their scaling relative to the values chosen by Hodgkin and Huxley (1952). Thus, for instance, $<\bar{g}_{\text{Na}}>$ = 1.2 stands for $\bar{g}_{\text{Na}}$ = 144 mS/cm$^{2}$ (i.e., x1.2 the value chosen by Hodgkin and Huxley; see Figure~\ref{fig:1}). Transition rates are similarly scaled by multiplication. Thus, for instance, the expression $<\beta_{n}(v)>$ = 0.75 stands for $0.75 \beta_{n}(v)$. The shaded areas added to the graphs of Figure~\ref{fig:1} suggest that such a linear scaling of transition rate functions is justified, as it captures most of the underlying variance. 

\begin{figure}[]
\centering
\includegraphics[width=12cm,height=9cm]{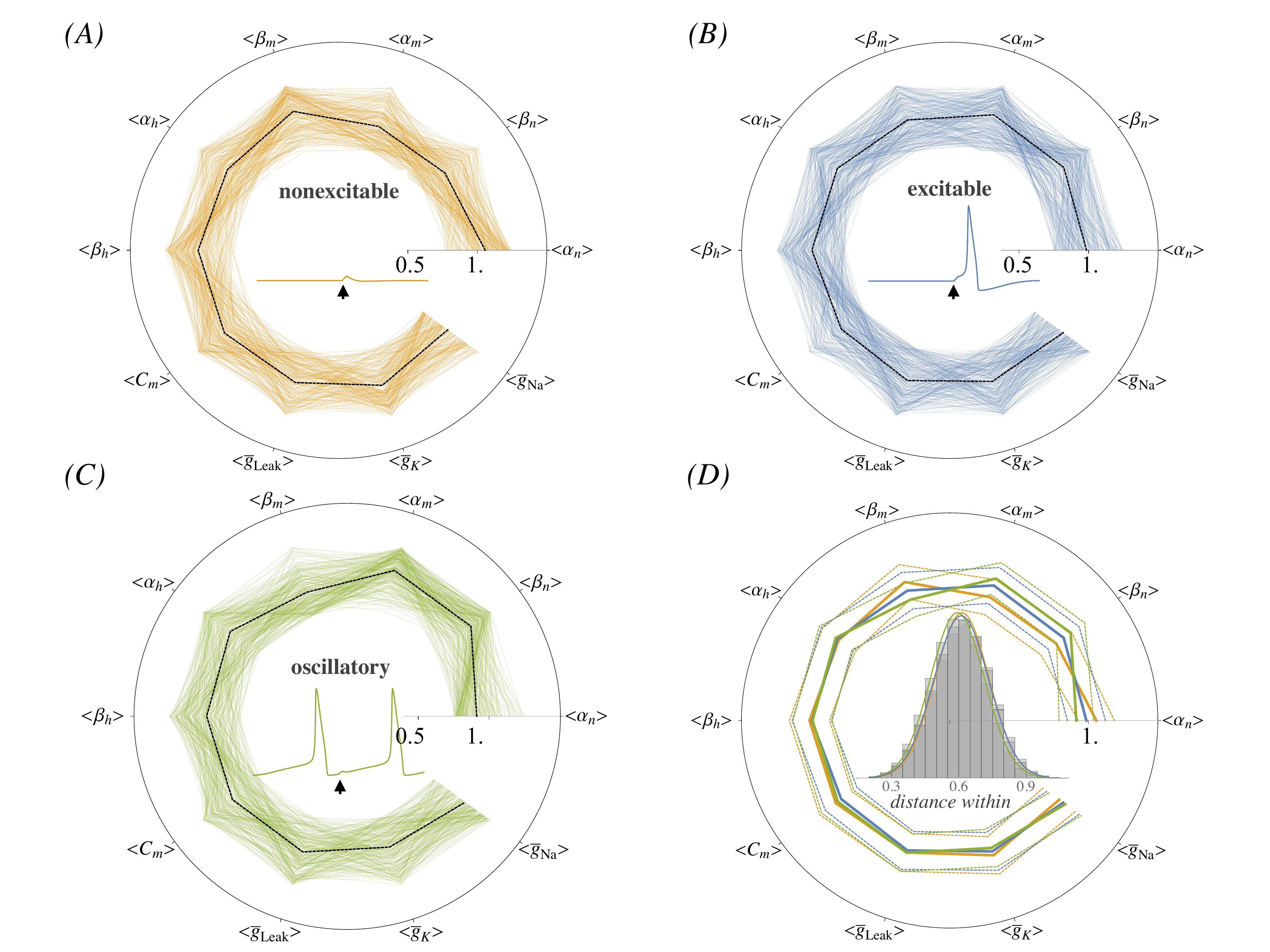}
\caption{\protect\rule{0ex}{20ex}Realizations (10,000) of a full Hodgkin-Huxley model; each realization is uniquely defined by a vector of ten parameters, expressed in terms of their scaling relative to the values chosen by Hodgkin and Huxley. Responses are classified to three excitability statuses (different colors): excitable (2225), nonexcitable (4884) and oscillatory (2891). Subsets of the results (200 for each excitability status) are presented in polar plots: Given a list of ten scaling parameters, the value of each parameter is depicted along its own (angular) axis, and the entire vector is depicted as a line that connects the ten scaling parameters. The standard Hodgkin-Huxley model would be a line passing through 1 for all scaling parameters. Mean vectors are depicted by dashed lines. The histograms in the bottom-right panel depict Euclidean distance between vectors of scaled parameters \textit{within} each of the three excitability classes.}
\label{fig:2}
\end{figure}

Hence, a realization of a Hodgkin-Huxley model is defined by a list of ten scaling parameters: {$<\alpha_{n}(v)>$, $<\beta_{n}(v)>$, $<\alpha_{m}(v)>$, $<\beta_{m}(v)>$, $<\alpha_{h}(v)>$, $<\beta_{h}(v)>$, $<C_{\text{m}}>$, $<\bar{g}_{\text{leak}}>$, $<\bar{g}_{\text{K}}>$, $<\bar{g}_{\text{Na}}>$}.  Assuming independence of the ten parameters, we randomly generated 10,000 such lists of scaling parameters with values range [0.75, 1.25], and numerically instantiated, each one of them, in a full Hodgkin-Huxley model. The resulting behaviors may be classified as nonexcitable (i.e. passive), excitable (i.e. a membrane that generates one spike in response to a short above-threshold stimulus), and oscillatory (i.e. pace-making). Resting membrane potentials of the excitable and non-excitable outcomes did not differ much over the $\pm$25\% deviation from the original Hodgkin and Huxley parameters, being -64.5$\pm$1.4 mV and -66.2$\pm$1.5 mV (respectively). 

To promote effective visualization of the ten-dimensional space, results are presented in the form of polar plots (Figure~\ref{fig:2}): Given a vector of ten scaling parameters, the value of each parameter is depicted along its own (angular) axis, and the entire vector is depicted as one line that connects the ten scaling parameters. The standard Hodgkin-Huxley model would be a line passing through 1 for all scaling parameters. Three separate polar plots (panels A, B and C) show that practically all three classes of excitability status (depicted by three different colors) are distributed throughout the 10-dimensional parameter space. Mean vectors in each of these cases are depicted by black dashed lines. For comparison, these three mean vectors and their corresponding standard deviations are plotted together in the polar plot of panel D. The Euclidean distance between the mean vector of excitable and the mean vectors of the other two solutions (non-excitable or oscillatory) is ca. 0.15, similar to the standard deviation of distances within each of them (inset to panel D). In other words, assuming complete independence of the parameters within the $\pm$25\% range of parametric variation, almost any randomly chosen vector of Hodgkin-Huxley parameters, regardless of its outcome (non-excitable, excitable or oscillatory), may be `pushed' to display any other excitability status by a minor manipulation of parameters.

\subsubsection*{Lower dimension Hodgkin-Huxley parameter space}

\begin{figure}[]
\centering
\includegraphics[width=12cm,height=9cm]{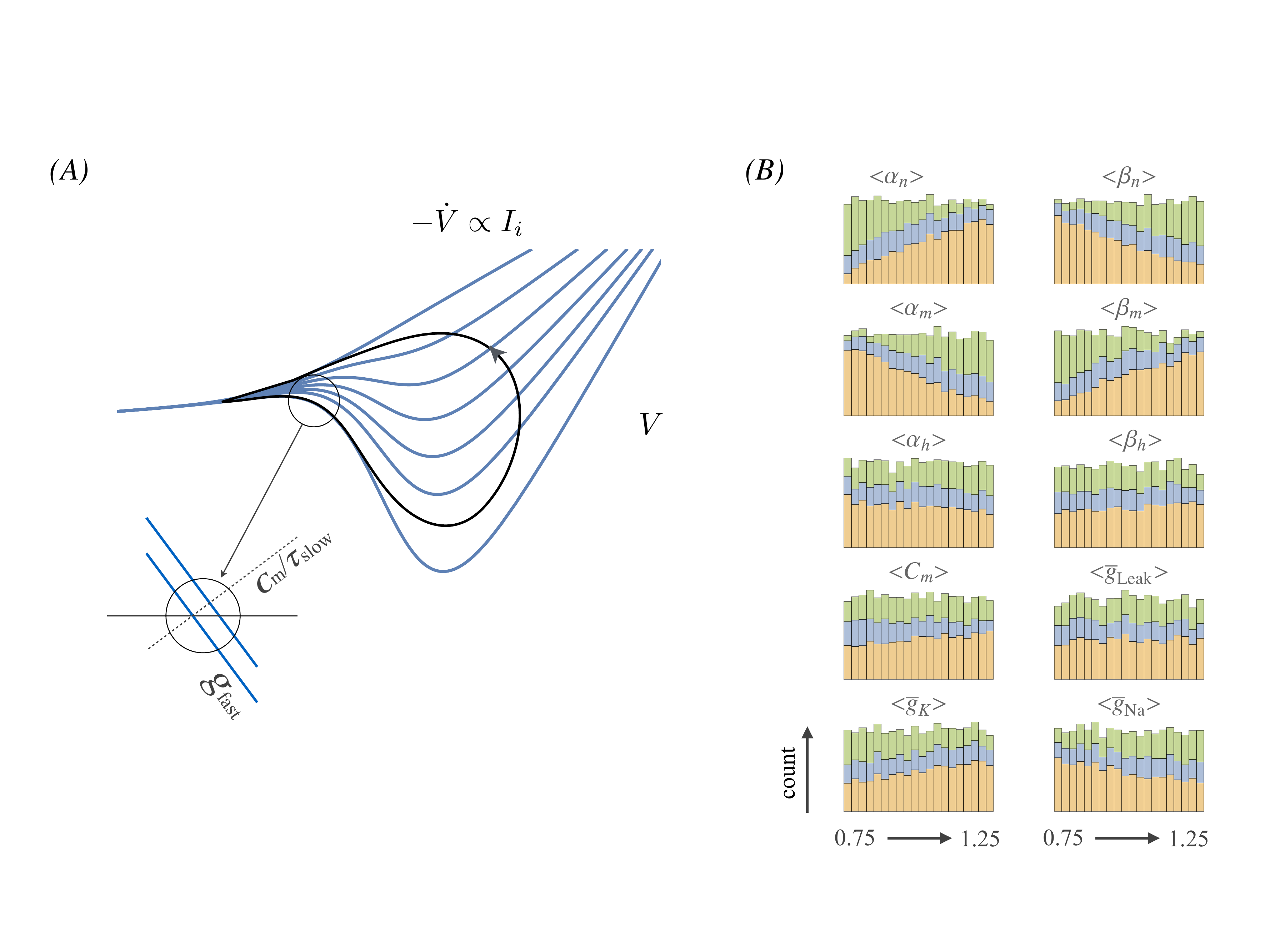}
\caption{\protect\rule{0ex}{20ex}(A) Idealized momentary current-voltage relations at different ratios of available sodium and potassium conductances (modified versions of Figures 11.9 \& 11.10 in Jack, Noble and Tsien, 1975). In different phases of the action potential, different momentary current-voltage relations determine the dynamics. The black continuous line depicts the resulting current-voltage trajectory during an action potential. Inset: a magnified version of the area at threshold (indicated in the main figure by a circle), about which the system is linearized. (B) histograms of the three excitability statuses, constructed from the data of Figure \ref{fig:2} (10000 Hodgkin-Huxley realizations), for each of the scaling parameters. Note that all parameters are freely fluctuating, simultaneously, over $\pm$25\%. }
\label{fig:3ab}
\end{figure}

We focus on the conditions for transition between excitable and non-excitable statuses. Several schematic momentary current-voltage relations of excitable membranes, during an action potential, are plotted in Figure~\ref{fig:3ab}A. Grossly speaking, the lower curve depicts current-voltage relations sampled by voltage-clamp steps from deeply hyperpolarized holding potential. The upper curve depicts current-voltage relations sampled by voltage-clamp steps from a relatively depolarized holding potential. During an action potential, where membrane voltage is a dynamical free variable, current-voltage relations slowly shift between these two extremes due to an evolving voltage-dependent restoring force, mediated by the opening of potassium channels and inactivation of sodium channels. The slow change in restoring force gives rise to a current-voltage closed trajectory depicted in Figure~\ref{fig:3ab}A (black continuous line). 

The general differential equation of the system is $C_{\text{m}} d^{2}V/dt^{2} + dI_{i}/dt = 0$. As pointed out by Jack, Noble and Tsien \cite{Jack1975} (Ch. 11), linearization about the threshold potential (inset to Figure~\ref{fig:3ab}A) leads to an expression of $dI_{i}/dt$ in terms of the momentary conductance ($-g_{\text{fast}}$) at threshold and the time constant ($\tau_{\text{slow}}$) for evolving restoring force. The condition for instability near threshold, one of the solutions to this equation (expressed in conductance units), is $-g_{\text{fast}}>C_{\text{m}}/\tau_{\text{slow}}$. Note that these lumped entities ($-g_{\text{fast}}$, $C_{\text{m}}$ and $\tau_{\text{slow}}$) may naturally be classified into the above groups of structural and kinetic parameters: $C_{\text{m}}$ is obviously structural; likewise, the fast conductance ($-g_{\text{fast}}$) that depends on the relative contribution of maximal sodium conductance. In contrast, the time scale for introduction of restoring force ($\tau_{\text{slow}}$) is a kinetic parameter because it depends on the actual transition rate functions governing the gating of sodium and potassium channels. 

Inspired by the above and related mathematical reductions of excitability \cite{Abbott1990,FitzHugh1961,Izhikevich2003}, we turned to the data of Figure~\ref{fig:2} in search for these two dimensions, expressed in terms of Hodgkin-Huxley parameters. To this aim, we constructed count histograms of the three excitability statuses for each of the scaling parameters (Figure~\ref{fig:3ab}B). These histograms show that the most critical determinants of excitability status are the rates of opening and closure of sodium and potassium channels ($\alpha_{m}(v)$, $\beta_{m}(v)$, $\alpha_{n}(v)$ and $\beta_{n}(v)$) and the maximal conductance of the membrane to the two ions ($\bar{g}_{\text{Na}}$, $\bar{g}_{\text{K}}$).\footnote{Similar results are obtained using the two ``far apart'' excitability statuses -- nonexcitable and oscillatory -- to extract principal components; not shown.} Distribution of excitability status is significantly less sensitive to transition rates involved in sodium conductance inactivation, as well as leak conductance. Membrane capacitance seems to have some effect. A possible interpretation of the histograms of Figure~\ref{fig:3ab}B is that, at least as a first approximation, $-g_{\text{fast}}$ may be assumed to be proportional to a ratio of structural parameters \textit{S} = $<\bar{g}_{\text{Na}}>/(<\bar{g}_{\text{Na}}> + <\bar{g}_{\text{K}}>$), whereas $\tau_{\text{slow}}$ may be assumed to be proportional to the ratio of kinetic parameters \textit{K} = $(<\alpha_{n}(v)>+<\beta_{m}(v)>)/(<\alpha_{n}(v)>+<\beta_{m}(v)>+<\alpha_{m}(v)>+<\beta_{n}(v)>)$. 

\begin{figure}[]
\centering
\includegraphics[width=12cm,height=9cm]{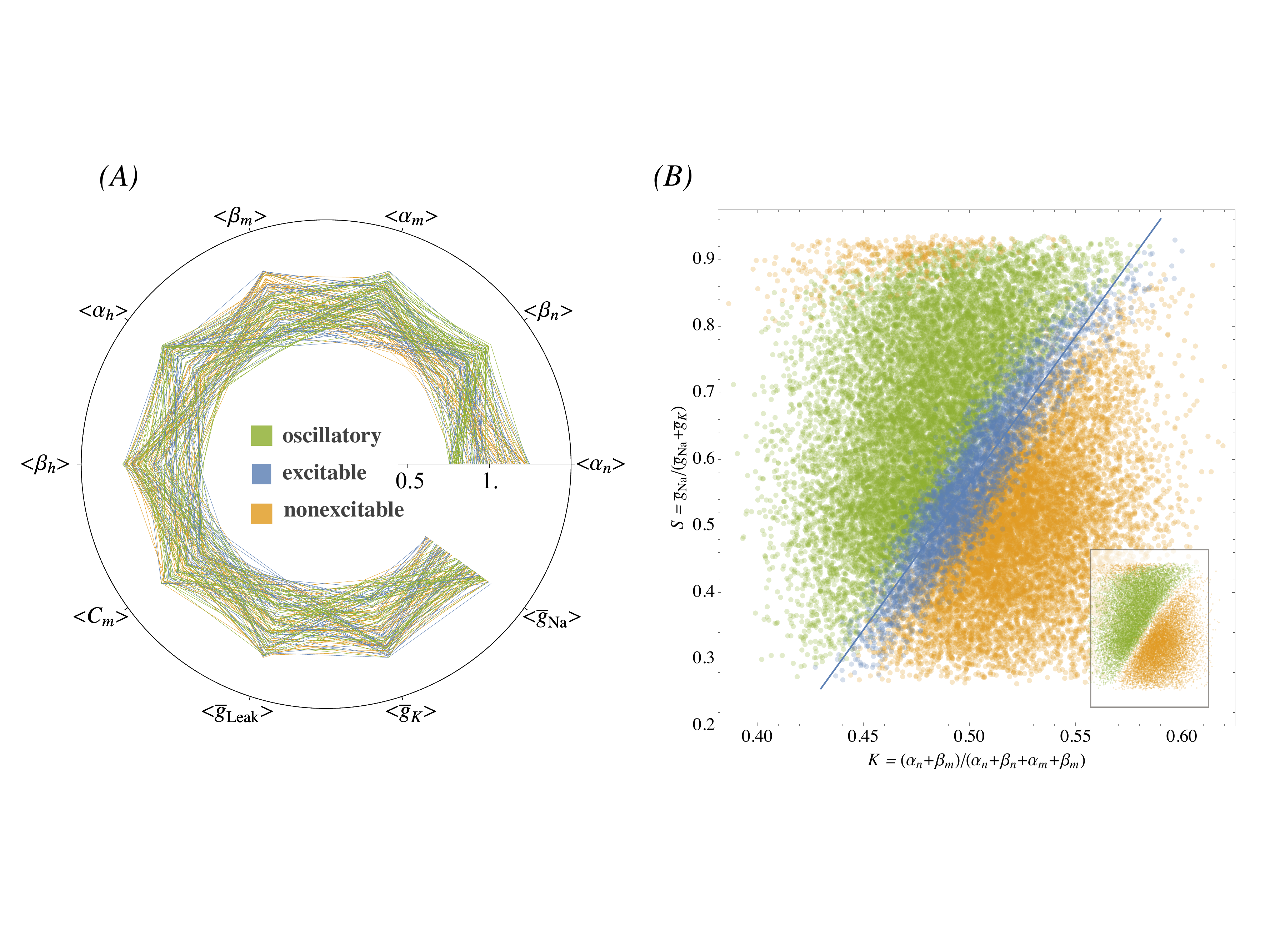}
\caption{\protect\rule{0ex}{20ex}(A) Subsets of 100 realizations from each of the three excitability statuses of Figure \ref{fig:2} are plotted together. (B) Realizations (30,000) of a full Hodgkin-Huxley model, covering parametric variations over the entire range indicated by Hodgkin and Huxley (see Figure \ref{fig:1}), classified (different colors) to three excitability statuses: excitable (4660), not excitable (12271) and oscillatory (13069). Linear regression through the excitable status cloud (blue) is depicted by a line, the equation of which is $S = 4.4K-1.6$. Inset: same plot with excitable points omitted. }
\label{fig:4ab}
\end{figure}

The murky enmeshment of solutions shown in Figure~\ref{fig:2} (a sample of which is re-plotted in Figure~\ref{fig:4ab}A, superposed) is significantly clarified when the data are arranged according to the values of \textit{S} and \textit{K} (Figure~\ref{fig:4ab}B): The three different excitability regimes are nicely clustered in three clouds. The oscillatory and nonexcitable phases are well-separated in the \textit{S--K} plane (inset to Figure~\ref{fig:4ab}B), whereas the borders separating the excitable phase from these other two are `soft' rather than sharp. The nonexcitable cloud in the upper left corner is due to excessive sodium conductance that stabilizes the membrane at a depolarized potential. 

\begin{figure}
\centering
\includegraphics[width=1\linewidth]{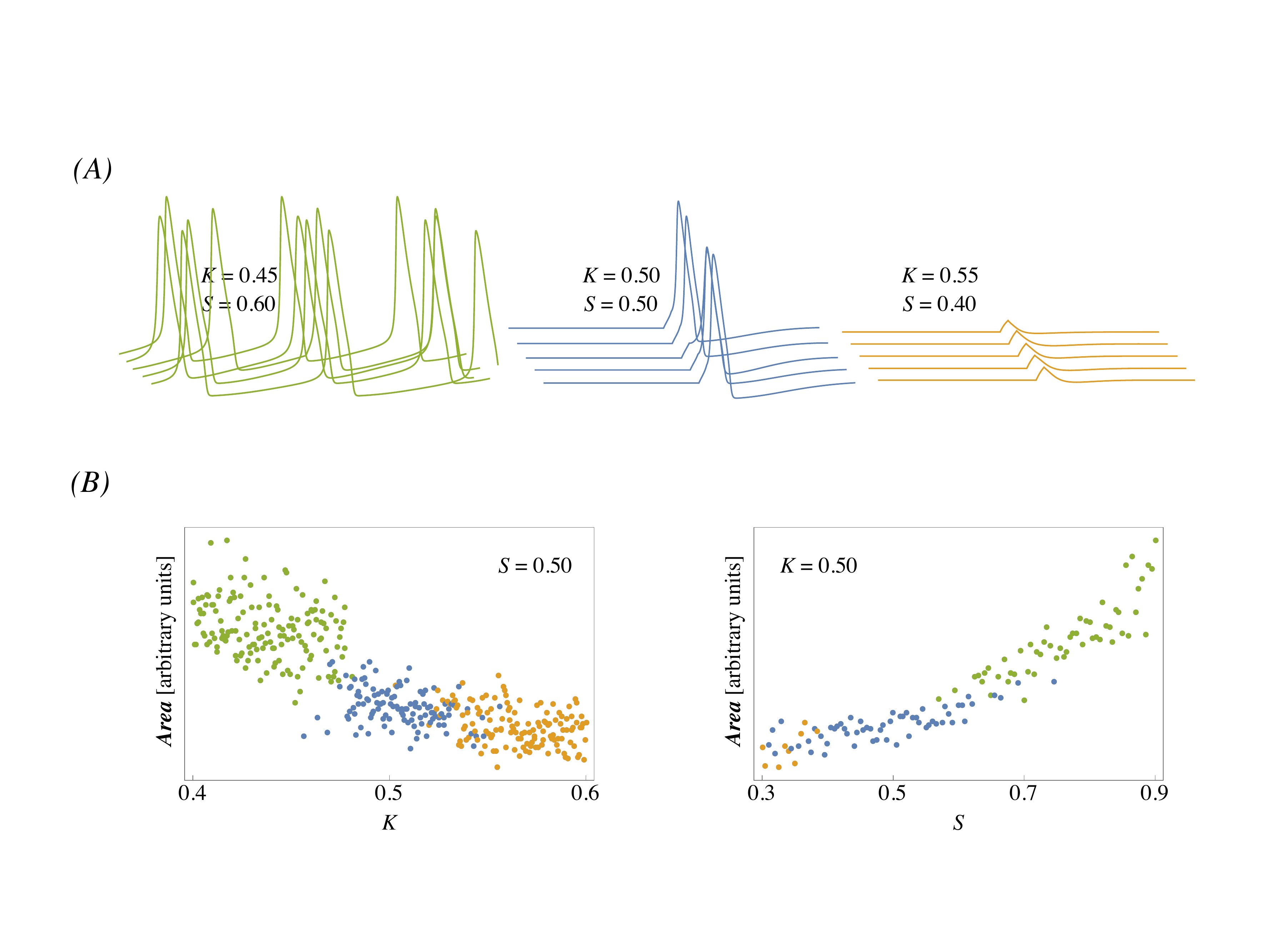}
\caption{(A) Five different instantiations for each of three S;K pairs (depicted within panel \ref{fig:4ab}B); stimulation amplitude 14 $\mu$A. (B) The integral of voltage response emitted during a simulated trace is sensitive to the position within the \textit{S--K} plain. The integral, calculated by summing the voltage values of all data points along the trace, relative to -65 mV, is presented in arbitrary units: \textit{left} -- gradual change of $K$ at $S = 0.5$, \textit{right} --  gradual change of $S$ at $K = 0.5$. Point color depicts excitability class. }
\label{fig:5ab}
\end{figure}

Note that in Figure~\ref{fig:4ab}B, four parameters (two sodium inactivation rates, capacitance and leak conductance) are not taken into account even though they are allowed to freely fluctuate. And yet, when examined in the two-dimensional \textit{S--K} space, the Hodgkin-Huxley model reveals order that is literally impossible to detect in the more explicit, higher dimensional representation of Figure~\ref{fig:4ab}A. The effectiveness of the dimensionality reduction is further supported in Figure~\ref{fig:5ab}A, where the outcomes of multiple realizations of three different \textit{S;K} pairs are shown. Each of these three pre-defined \textit{S;K} pairs (0.60;0.45, 0.50;0.50, and 0.40;0.55) was realized 30 times by adjusting $<\alpha_{m}(v)>$ and $<\bar{g}_{\text{K}}>$ to the other four, randomly generated, Hodgkin-Huxley parameters ($<\alpha_{n}(v)>$, $<\beta_{m}(v)>$, $<\beta_{n}(v)>$ and $<\bar{g}_{\text{Na}}>$). Five of 30 are shown for each value; the rest are comparable. Clearly, the values of the lumped \textit{S} and \textit{K} dimensions are better predictions of the outcome, than the individual Hodgkin-Huxley parameters. 

\begin{figure}
\centering
\includegraphics[width=1\linewidth]{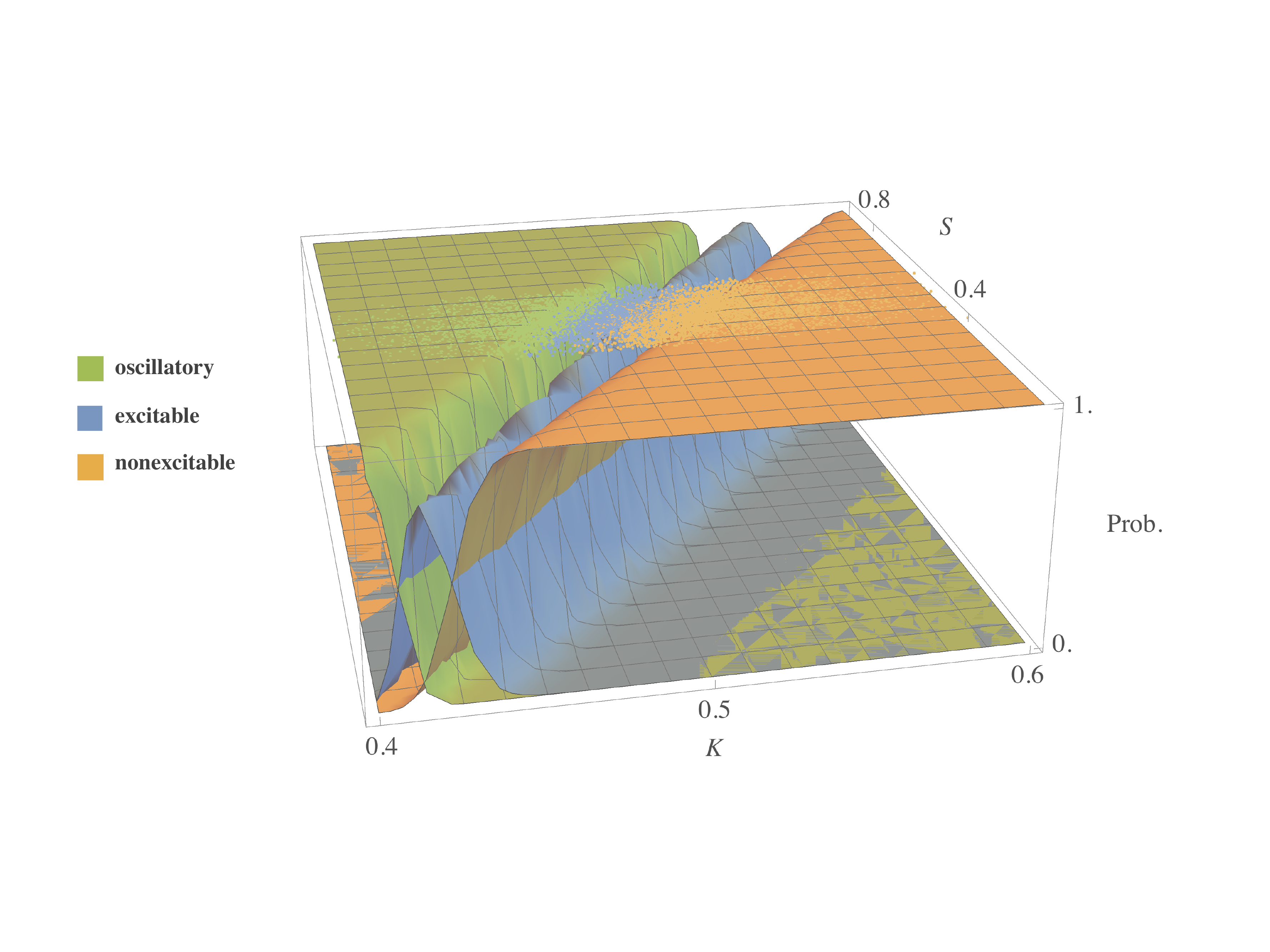}
\caption{Classification of excitability statuses (data of Figure 2) using a linear kernel support vector machine (SVM); 80\% training set. Each surface (oscillatory, excitable, nonexcitable; color coded) represents the probability (z-axis) of a given combination of $K$ and $S$ to give rise to its corresponding excitability status. Thus, for instance, the probability of a point $K=0.42$ and $S=0.7$ to yield an oscillatory (depicted green) excitability status is practically 1. The colored points at the top of the image are the actual data points, similar to the form of presentation used in Figure 4B. }
\label{fig:6}
\end{figure}

Several points deserve attention in relation to the numerically calculated \textit{S--K} plane of Figure~\ref{fig:4ab}B. First, the borders between the three phases are fairly steep (note different ranges of \textit{S} and \textit{K} axes). The immediate implication of this steepness is that in its two extreme statuses (nonexcitable and oscillatory), the system is relatively immune to variations in maximal conductances. Stated differently, it is sufficient to use ionic channels that set the \textit{K} dimension below (ca.) 0.45 to obtain a pace-maker that is insensitive to fluctuations in density of channel proteins; the system maintains its pace-making nature over factor 3 in the value of the structural ($S$) dimension. The same can be said about nonexcitable membranes: setting the \textit{K} dimension above (ca.) 0.55 to obtain a nonexcitable system that is insensitive to fluctuations in density of channel proteins. Second, the \textit{K} dimension is a rational function of first degree, a simple combination of concrete Hodgkin-Huxley kinetic parameters; as such it buffers the effect of changes in individual rates. This, one might expect, would also be the case for the \textit{S} dimension where more than two voltage-dependent conductances are involved. Third -- moving within the \textit{S--K} plane has an interpretable effect on the response shape (Figure~\ref{fig:5ab}B): the integral of voltage response emitted during a simulated trace is sensitive to the position within the \textit{S--K} plane. Naturally, difference between points in the nonexcitable phase is very small, if at all. Fourth -- a seemingly technical point but of potential interest: given scaled Hodgkin-Huxley parameters, one can calculate the resulting excitability status without resorting to simulation. This means that the non-linearity of the model does not change the behavior actually predicted from a low-dimensional representation of the system \cite{sobie2009parameter}. And, fifth -- admittedly, our theoretically inspired choice of the rational functions that express \textit{S} and \textit{K} is one of many possible interpretations to relations between scaling parameters. To further justify this choice, we submitted the whole data set to a linear support vector machine (SVM) algorithm. The results are presented in Figure~\ref{fig:6}, where the test set and the probability distribution of each class are plotted as a function of \textit{S} and \textit{K}. The accuracy of classifying the outcome of a full Hodgkin-Huxley model, based on the values of \textit{S} and \textit{K} is 0.89, suggests that our theoretically-inspired reduction of the Hodgkin-Huxley model to an \textit{S--K} plane is judicious. It remains to be seen whether the dimensionality reduction approach we used above can be applied to models with many more channel types. 

\subsubsection*{Closed-loop control of excitability in the \textit{S--K} plane: the case of sodium conductance slow inactivation}

Being embedded in \textit{S--K}, the seemingly complicated and parameter sensitive system becomes tractable, enabling regulation by an activity-dependent rule acting on one physiological entity. A most straight-forward regulation rule would involve inverse relations between electrical activity (say, integral of membrane potential depolarizations) and the effective or actual value of the structural dimension (\textit{S}). Many physiological processes that modulate membrane ion channels may realize such adaptation, covering a wide range of spatial and temporal scales \cite{Marom2010}. For instance: (1) Slow inactivation of sodium conductance, which is a local modulatory mechanism that operates over seconds to many minute time scales \cite{Catterall2015,Marom2016,Ruben1992,Silva2014,Toib1998,Ulbricht2005,Vilin2001}; or, (2) calcium-dependent activation of potassium conductance, a local and relatively fast time scale mechanism \cite{Brenner2000,Ghatta2006,Sah1996}, or (3) regulation of sodium and/or potassium channel protein expression, an arguably global but definitely slow time scale mechanism \cite{Bucher2011,Schulz2007}. Each of these in itself naturally constrains the system to hover about the excitable phase by pushing the value of \textit{S} downward when above the diagonal, or upwards when below. Other regulatory mechanisms may be envisioned, implementing (for instance) temporal integration of subthreshold activity by slow inactivation of potassium channels \cite{Marom1994,Storm1988}, or maintenance of pace-making activity by regulation of IK$_{\text{\textit{f}}}$ conductance \cite{Baruscotti2005}.

Of the above-mentioned spectrum of physiological modulatory mechanisms, slow inactivation of sodium channels is especially interesting. While acknowledged from the early days of membrane electrophysiology [reviewed in \cite{Ulbricht2005}], the concept of slow inactivation of sodium channels as a means to maintain excitability status amidst parametric variations, has remained relatively neglected. What makes slow inactivation a powerful regulatory mechanism is its impacts on the \textit{effective} value of $\bar{g}_{\text{Na}}$, covering a range of time scales \cite{Ellerkmann:2001bs,fleidervish1996slow,Silva2014,Toib1998,Ulbricht2005,Vilin2001}. Furthermore, inactivation is `local' in the sense that it does not require central control; it occurs automatically as a consequence of activity. Thus, one might picture it acting as a distributed normalizing force in extended excitable tissues (e.g., long axons or electrically coupled excitable cells). 


\begin{figure}[]
\centering
\includegraphics[width=12cm,height=9cm]{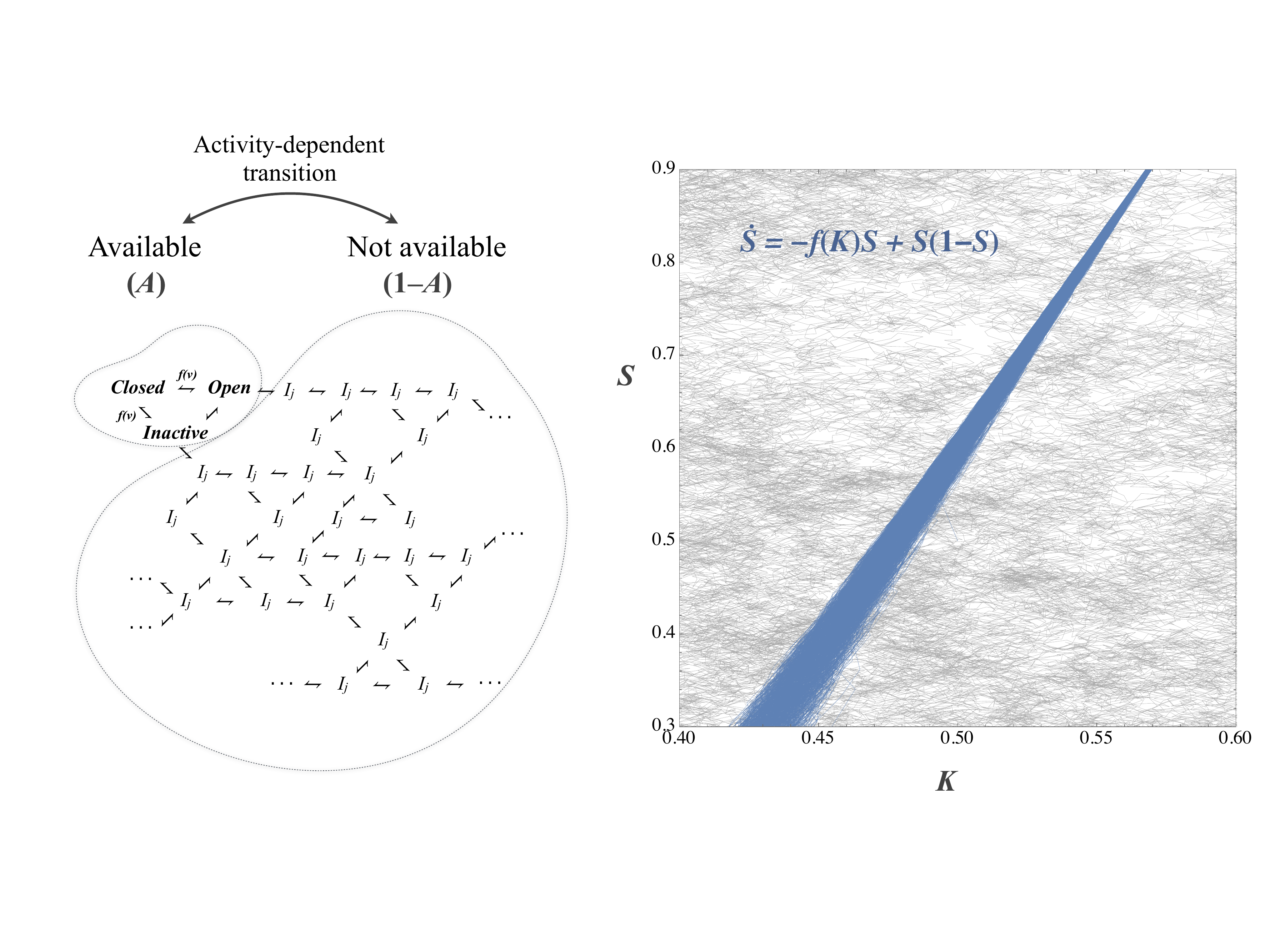}
\caption{\protect\rule{0ex}{20ex}Left: a schematic representation of sodium channel states, with many slow inactivation states. Right: Demonstration of maintenance of excitability in $S--K$ plane given parametric variation, controlled by activity-dependent transitions of sodium channels between available and not-available sets of states. The simulation describes a 200000 steps random walk process, beginning at 0.5;0.5 ($S;K$). The gray trajectory (squiggles-like) depicts a random walk where $K_{n+1} = K_{n} \pm0.01$ and $S_{n+1} = S_{n}\pm0.01$. The blue line depicts a walk where $K_{n+1} = K_{n} \pm0.01$ and $S_{n+1} = S_{n} \pm0.01 - (2.6-4.4 K) S + S (1-S)$.  The slope corresponds to the fitted function of Figure \ref{fig:4ab}B. }
\label{fig:7}
\end{figure}

To demonstrate the potential impacts of slow inactivation on dynamics within the \textit{S--K} plane, we focus on channel gating beyond the time scale of a single action potential. Slow inactivation is represented as a macroscopic system, where channels move -- in an activity-dependent manner -- between two states: available and not-available (Figure~\ref{fig:7}, left panel), depicted $A\leftrightarrow(1-A)$. The first (\textit{A}) is the set of states that includes, besides the open state itself, all the states from which the channel may arrive to the open state within the time scale of a single action potential, i.e. -- the closed states and the very first inactive states that are treated in standard Hodgkin and Huxley formalism for the action potential generation. In other words, a channel in \textit{A} is available for conducting ions within the time scale of a single action potential. In contrast, $(1-A)$ is a large, interconnected pool of slow inactive states (depicted \textit{I}j's in the scheme of Figure \ref{fig:7}) from which transition to the open state within the time scale of an action potential is impossible. Recent structural analyses suggest that the large space of slow inactive states $(1-A)$ might reflect the many distorted versions of the functional protein under conditions where the organization, otherwise enforced by hyperpolarized membrane potential, is compromised upon extensive depolarizing activity \cite{Catterall2015}. Theory and experiments \cite{Ellerkmann:2001bs,Goychuk:2004kx,Marom2016,Marom2010,millhauser1988diffusion,millhauser1988rate,Toib1998} show that in such a scheme, the multiplicity of slow inactive states entails a power-law scaling of recovery from $(1-A)$ to $A$ as a function of time spent in $(1-A)$. This implies a potential to become dormant in an activity-dependent manner for a duration ranging from tens of milliseconds to many minutes and possibly hours. Thus, unlike standard Hodgkin-Huxley gates, the rate of recovery from slow inactivation does not have a uniquely defined characteristic time scale. Rather, the time scale is determined by the distribution of channels in the space of inactive states, which, in turn, is dictated by the history of activation. The kinetics of $A\leftrightarrow(1-A)$ may be qualitatively described by an adaptive rate model \cite{Gal2013,excitability2014single,Marom2010,marom2009adaptive,xu2017dynamical}, a logistic-like function of the form: $\dot{A} = -f(\gamma)A + g(A)(1-A)$, where $f$ is a function of some general activity measure $\gamma$, and $g(A)$ is a monotonically increasing function of the system state $A$. The model gives rise to a wide range of time scales of recovery from inactivation and assures a non-zero stable point at $(1-\gamma)$, on the edge between excitable and nonexcitable \cite{Gal2013}. 

Mapping the above picture to the terms used in the present work, it is instructive to think of $A$ as $<\bar{g}_{\text{Na}}>$, i.e. a scaling parameter of maximal sodium conductance. In the original Hodgkin and Huxley formalism, $\bar{g}_{\text{Na}}$ is a structural constant that sets limits on the instantaneous (at the scale of milliseconds) input--output relations of the membrane. But when long-term effects are sought, $<\bar{g}_{\text{Na}}>$ might be treated as a dynamic variable that modulates residence in a reservoir of slow-inactivation configurations, pulling channels away from the system as a function of activity. Note that where $<\bar{g}_{\text{K}}>$ is constant and where $<\bar{g}_{\text{Na}}>$ = $A$, the adaptive rate model qualitatively captures the dynamics of $<\bar{g}_{\text{Na}}>$/($<\bar{g}_{\text{Na}}>$+$<\bar{g}_{\text{K}}>$). Indeed, the right panel of Figure~\ref{fig:7} demonstrates that application of a simple adaptive rate model $\dot{S} = -f(K)S + f(S)(1-S)$, where $f(\gamma)$ is substituted by $f(K)$, reveals the potential of sodium conductance slow inactivation to maintain excitability amid parametric variations. Slow inactivation restrains the system to a diagonal in the \textit{S--K} plane: the blue line depicts a case where the kinetic dimension ($K$) walks randomly while $\dot{S}$ follows an adaptive rate formalism; a gray line that connects between points (looks like squiggles) depicts the path of excitability status in a control condition, where both the kinetic dimension ($K$) and the structural dimension ($S$) walk randomly. 

\subsubsection*{Concluding remarks}

In an era marked by capacity to collect data at ever-increasing resolution, it is important to identify proper scales in analyses of cellular phenomena, 
scales that matter to the system \cite{Transtrum2015}, scales where the phenomenon of interest is low-dimensional, regulatable by simple physiological processes and explainable in simple physiological terms. What makes membrane excitability -- a fundamental physiological phenomenon -- particularly attractive to study in this context, is the existence of a relatively sound theory in general, its application in the Hodgkin-Huxley formalism in particular, and its amenability to experimental manipulations at both microscopic (channel protein) and macroscopic (membrane potential) levels. 

Practically all of the homeostatic-based models of excitability regulation used in the past have kept kinetics constant and only looked at channel densities \cite{lemasson1993activity,OLeary2014,O_Leary_2018}. The present study suggests that when the problem is examined in a lower dimension, a simple control rule that relies on slow inactivation -- a ubiquitous protein intrinsic process -- can deal with fluctuations in both structural and kinetic parameters. This homeostatic mechanism is local, independent of protein synthesis and operates over a wide range of time scales (milliseconds to many minutes). 


We speculate that activity dependence of protein kinetics at relatively slow time scales, entailed by multiplicity of protein states, is a general ``automatic'' and local means for stabilization of cellular function.

\subsubsection*{Methods}All the simulations and analyses were performed within \textit{Mathematica} (Wolfram Research, Inc.) environment. Data of Figures 2, 3, 4 and 5 were generated using Hodgkin-Huxley equations as they appear in the original manuscript (1952). The duration of each simulation epoch was 90 msec, including an initial 50 msec relaxation phase. Stimulus (7 $\mu$A, 1 msec) was delivered 70 msec into the epoch. A sorting algorithm for excitability status (excitable, non-excitable or oscillatory) was constructed, which is based on the time and the number of spikes following the relaxation phase. The sorting algorithm was validated by eye-inspection of multiple sets of 300 sorted epochs. To generate Figure 7, a support vector machine (SVM) algorithm with linear kernel was implemented within \textit{Mathematica}, using an 80\% training set.

\subsubsection*{Acknowledgments}This work was partially supported by research grants from the Leir Foundation (EM, SM) the National Institutes of Health (EM), and the Israel Science Foundation (SM). The authors thank Omri Barak, Asaf Gal and Daniel Soudry for helpful comments.

\bibliographystyle{unsrt} 
\bibliography{EmEx1}

\begin{thebibliography}{10}

\bibitem{Hodgkin1952}
A~L Hodgkin and A~F Huxley.
\newblock {A quantitative description of membrane current and its application
  to conduction and excitation in nerve}.
\newblock {\em J. Physiol.}, 117:500--544, 1952.

\bibitem{Johnstone2016}
Ross~H Johnstone, Eugene T~Y Chang, R{\'{e}}mi Bardenet, Teun~P {De Boer},
  David~J Gavaghan, Pras Pathmanathan, Richard~H Clayton, and Gary~R Mirams.
\newblock {Uncertainty and variability in models of the cardiac action
  potential: Can we build trustworthy models?}
\newblock {\em Journal of molecular and cellular cardiology}, 96:49--62, 2016.

\bibitem{OLeary2015}
Timothy O'Leary, Alexander~C Sutton, and Eve Marder.
\newblock {Computational models in the age of large datasets}.
\newblock {\em Current opinion in neurobiology}, 32:87--94, 2015.

\bibitem{mirams2016uncertainty}
Gary~R Mirams, Pras Pathmanathan, Richard~A Gray, Peter Challenor, and
  Richard~H Clayton.
\newblock Uncertainty and variability in computational and mathematical models
  of cardiac physiology.
\newblock {\em The Journal of physiology}, 594(23):6833--6847, 2016.

\bibitem{Golowasch2002}
Jorge Golowasch, Mark~S Goldman, L~F Abbott, and Eve Marder.
\newblock {Failure of averaging in the construction of a conductance-based
  neuron model}.
\newblock {\em Journal of Neurophysiology}, 87(2):1129--1131, 2002.

\bibitem{Marder2011}
Eve Marder.
\newblock {Variability, compensation, and modulation in neurons and circuits}.
\newblock {\em Proceedings of the National Academy of Sciences}, 108(Supplement
  3):15542--15548, 2011.

\bibitem{Brenner2015}
Naama Brenner, Erez Braun, Anna Yoney, Lee Susman, James Rotella, and Hanna
  Salman.
\newblock {Single-cell protein dynamics reproduce universal fluctuations in
  cell populations}.
\newblock {\em The European Physical Journal E}, 9(38):1--9, 2015.

\bibitem{AsafGal12012010}
Asaf Gal, Danny Eytan, Avner Wallach, Maya Sandler, Jackie Schiller, and Shimon
  Marom.
\newblock {Dynamics of Excitability over Extended Timescales in Cultured
  Cortical Neurons}.
\newblock {\em J. Neurosci.}, 30(48):16332--16342, 2010.

\bibitem{Li2011}
Gene-Wei Li and X~Sunney Xie.
\newblock {Central dogma at the single-molecule level in living cells}.
\newblock {\em Nature}, 475(7356):308, 2011.

\bibitem{Marder2014}
Eve Marder, Timothy O'Leary, and Sonal Shruti.
\newblock {Neuromodulation of circuits with variable parameters: single neurons
  and small circuits reveal principles of state-dependent and robust
  neuromodulation}.
\newblock {\em Annual Review of Neuroscience}, 37:329--346, 2014.

\bibitem{OLeary2014}
Timothy O'Leary, Alex~H Williams, Alessio Franci, and Eve Marder.
\newblock {Cell types, network homeostasis, and pathological compensation from
  a biologically plausible ion channel expression model}.
\newblock {\em Neuron}, 82(4):809--821, 2014.

\bibitem{Raj2008}
Arjun Raj and Alexander van Oudenaarden.
\newblock {Nature, nurture, or chance: stochastic gene expression and its
  consequences}.
\newblock {\em Cell}, 135(2):216--226, 2008.

\bibitem{Sigal2006}
Alex Sigal, Ron Milo, Ariel Cohen, Naama Geva-Zatorsky, Yael Klein, Yuvalal
  Liron, Nitzan Rosenfeld, Tamar Danon, Natalie Perzov, and Uri Alon.
\newblock {Variability and memory of protein levels in human cells}.
\newblock {\em Nature}, 444(7119):643, 2006.

\bibitem{Marom2016}
S~Marom.
\newblock {Emergence and maintenance of excitability: kinetics over structure}.
\newblock {\em Current opinion in neurobiology}, 40:66--71, 2016.

\bibitem{Daly2015}
Aidan~C Daly, David~J Gavaghan, Chris Holmes, and Jonathan Cooper.
\newblock {Hodgkin--Huxley revisited: reparametrization and identifiability
  analysis of the classic action potential model with approximate Bayesian
  methods}.
\newblock {\em Royal Society open science}, 2(12):150499, 2015.

\bibitem{sarkar2010regression}
Amrita~X Sarkar and Eric~A Sobie.
\newblock Regression analysis for constraining free parameters in
  electrophysiological models of cardiac cells.
\newblock {\em PLoS computational biology}, 6(9):e1000914, 2010.

\bibitem{sobie2009parameter}
Eric~A Sobie.
\newblock Parameter sensitivity analysis in electrophysiological models using
  multivariable regression.
\newblock {\em Biophysical journal}, 96(4):1264--1274, 2009.

\bibitem{Braun2015}
Erez Braun.
\newblock {The unforeseen challenge: from genotype-to-phenotype in cell
  populations}.
\newblock {\em Reports on Progress in Physics}, 78(3):36602, 2015.

\bibitem{lemasson1993activity}
G.~LeMasson, E.~Marder, and LF~Abbott.
\newblock Activity-dependent regulation of conductances in model neurons.
\newblock {\em Science}, 259:1915--1915, 1993.

\bibitem{liu1998model}
Zheng Liu, Jorge Golowasch, Eve Marder, and LF~Abbott.
\newblock A model neuron with activity-dependent conductances regulated by
  multiple calcium sensors.
\newblock {\em Journal of Neuroscience}, 18(7):2309--2320, 1998.

\bibitem{O_Leary_2018}
Timothy O'Leary.
\newblock Homeostasis, failure of homeostasis and degenerate ion channel
  regulation.
\newblock {\em Current Opinion in Physiology}, 2:129--138, Apr 2018.

\bibitem{Schulz2007}
David~J Schulz, Jean-Marc Goaillard, and Eve~E Marder.
\newblock {Quantitative expression profiling of identified neurons reveals
  cell-specific constraints on highly variable levels of gene expression}.
\newblock {\em Proceedings of the National Academy of Sciences},
  104(32):13187--13191, 2007.

\bibitem{OLeary2013}
Timothy O'Leary, Alex~H Williams, Jonathan~S Caplan, and Eve Marder.
\newblock {Correlations in ion channel expression emerge from homeostatic
  tuning rules}.
\newblock {\em Proceedings of the National Academy of Sciences},
  110(28):E2645--E2654, 2013.

\bibitem{Abbott1990}
L~F Abbott and Thomas~B Kepler.
\newblock {Model neurons: From Hudgkin-Huxley to Hopfield}.
\newblock In {\em Statistical mechanics of neural networks}, pages 5--18.
  Springer, 1990.

\bibitem{FitzHugh1961}
Richard FitzHugh.
\newblock {Impulses and physiological states in theoretical models of nerve
  membrane}.
\newblock {\em Biophysical journal}, 1(6):445--466, 1961.

\bibitem{Izhikevich2003}
Eugene~M Izhikevich.
\newblock {Simple model of spiking neurons}.
\newblock {\em IEEE Transactions on neural networks}, 14(6):1569--1572, 2003.

\bibitem{Jack1975}
James Julian~Bennett Jack, Denis Noble, and Richard~W Tsien.
\newblock {Electric current flow in excitable cells}.
\newblock 1975.

\bibitem{Marom2010}
S~Marom.
\newblock {Neural timescales or lack thereof}.
\newblock {\em Prog. Neurobiol.}, 90(1):16--28, 2010.

\bibitem{Catterall2015}
W~A Catterall and N~Zheng.
\newblock {Deciphering voltage-gated Na+ and Ca++ channels by studying
  prokaryotic ancestors}.
\newblock {\em Trends in Biochemical Sciences}, 40(9):526--534, 2015.

\bibitem{Ruben1992}
Peter~C Ruben, John~G Starkus, and MartinD Rayner.
\newblock {Steady-state availability of sodium channels. Interactions between
  activation and slow inactivation}.
\newblock {\em Biophysical journal}, 61(4):941--955, 1992.

\bibitem{Silva2014}
J~R Silva.
\newblock {Slow inactivation of Na{\^{}}+ channels}.
\newblock In Peter~C Ruben, editor, {\em Handbook of Experimental
  Pharmacology}, pages 33--49. Springer Berlin Heidelberg, 2014.

\bibitem{Toib1998}
A~Toib, V~Lyakhov, and S~Marom.
\newblock {Interaction between duration of activity and time course of recovery
  from slow inactivation in mammalian brain Na+ channels}.
\newblock {\em Journal of Neuroscience}, 18(5):1893--1903, 1998.

\bibitem{Ulbricht2005}
W~Ulbricht.
\newblock {Sodium channel inactivation: molecular determinants and modulation}.
\newblock {\em Physiol. Rev.}, 85(4):1271--1301, 2005.

\bibitem{Vilin2001}
Yuriy~Y Vilin and Peter~C Ruben.
\newblock {Slow inactivation in voltage-gated sodium channels}.
\newblock {\em Cell biochemistry and biophysics}, 35(2):171--190, 2001.

\bibitem{Brenner2000}
Robert Brenner, Tim~J Jegla, Alan Wickenden, Yi~Liu, and Richard~W Aldrich.
\newblock {Cloning and functional characterization of novel large conductance
  calcium-activated potassium channel $\beta$ subunits, hKCNMB3 and hKCNMB4}.
\newblock {\em Journal of Biological Chemistry}, 275(9):6453--6461, 2000.

\bibitem{Ghatta2006}
Srinivas Ghatta, Deepthi Nimmagadda, Xiaoping Xu, and Stephen~T O'Rourke.
\newblock {Large-conductance, calcium-activated potassium channels: structural
  and functional implications}.
\newblock {\em Pharmacology {\&} therapeutics}, 110(1):103--116, 2006.

\bibitem{Sah1996}
Pankaj Sah.
\newblock {Ca 2+-activated K+ currents in neurones: types, physiological roles
  and modulation}.
\newblock {\em Trends in neurosciences}, 19(4):150--154, 1996.

\bibitem{Bucher2011}
Dirk Bucher and Jean-Marc Goaillard.
\newblock {Beyond faithful conduction: short-term dynamics, neuromodulation,
  and long-term regulation of spike propagation in the axon}.
\newblock {\em Progress in neurobiology}, 94(4):307--346, 2011.

\bibitem{Marom1994}
S~Marom and L~F Abbott.
\newblock {Modeling state-dependent inactivation of membrane currents}.
\newblock {\em Biophysical journal}, 67(2):515--520, 1994.

\bibitem{Storm1988}
Johan~F Storm.
\newblock {Temporal integration by a slowly inactivating K+ current in
  hippocampal neurons}.
\newblock {\em Nature}, 336(6197):379--381, 1988.

\bibitem{Baruscotti2005}
Mirko Baruscotti, Annalisa Bucchi, and Dario DiFrancesco.
\newblock {Physiology and pharmacology of the cardiac pacemaker ("funny")
  current}.
\newblock {\em Pharmacology {\&} therapeutics}, 107(1):59--79, 2005.

\bibitem{Ellerkmann:2001bs}
R.~K. Ellerkmann, V.~Riazanski, C.~E. Elger, B.~W. Urban, and H.~Beck.
\newblock Slow recovery from inactivation regulates the availability of
  voltage-dependent sodium channels in hippocampal granule cells, hilar neurons
  and basket cells.
\newblock {\em J Physiol}, 532(Pt 2):385--97, 2001.

\bibitem{fleidervish1996slow}
Ilya~A Fleidervish, A~Friedman, and MJ~Gutnick.
\newblock Slow inactivation of na+ current and slow cumulative spike adaptation
  in mouse and guinea-pig neocortical neurones in slices.
\newblock {\em The Journal of Physiology}, 493(1):83--97, 1996.

\bibitem{Goychuk:2004kx}
I~Goychuk and P~Hanggi.
\newblock Fractional diffusion modeling of ion channel gating.
\newblock {\em Physical Review E}, 70(5), 2004.

\bibitem{millhauser1988diffusion}
G.L. Millhauser, E.E. Salpeter, and R.E. Oswald.
\newblock {Diffusion models of ion-channel gating and the origin of power-law
  distributions from single-channel recording}.
\newblock {\em Proceedings of the National Academy of Sciences},
  85(5):1503--1507, 1988.

\bibitem{millhauser1988rate}
GL~Millhauser, EE~Salpeter, and RE~Oswald.
\newblock {Rate-amplitude correlation from single-channel records. A hidden
  structure in ion channel gating kinetics?}
\newblock {\em Biophysical journal}, 54(6):1165, 1988.

\bibitem{Gal2013}
A~Gal and S~Marom.
\newblock {Self-organized criticality in single-neuron excitability}.
\newblock {\em Physical Review E}, 88(6):62717, 2013.

\bibitem{excitability2014single}
Asaf Gal and Shimon Marom.
\newblock Single neuron response fluctuations: A self-organized criticality
  point of view.
\newblock In Dietmar Plenz and Ernst Niebur, editors, {\em Criticality in
  Neural Systems}, chapter~11, pages 255--271. John Wiley \& Sons, 2014.

\bibitem{marom2009adaptive}
S.~Marom.
\newblock {Adaptive transition rates in excitable membranes}.
\newblock {\em Frontiers in Computational Neuroscience},
  3(2):doi:10.3389/neuro.10.002.2009, 2009.

\bibitem{xu2017dynamical}
Tie Xu and Omri Barak.
\newblock Dynamical timescale explains marginal stability in excitability
  dynamics.
\newblock {\em Journal of Neuroscience}, 37(17):4508--4524, 2017.

\bibitem{Transtrum2015}
M~K Transtrum, B~Machta, K~Brown, B~C Daniels, C~R Myers, and J~P Sethna.
\newblock {Perspective: Sloppiness and Emergent Theories in Physics, Biology,
  and Beyond}.
\newblock {\em J. Chem. Phys.}, 143(010901), 2015.

\end{thebibliography}

\end{document}